%% file: robust.tex
\documentclass[conference]{IEEEtran}

\usepackage{graphicx}
\usepackage[table,xcdraw]{xcolor}
\usepackage{algorithm}
\usepackage{algpseudocode}
\usepackage{bm}
\usepackage{amsmath}
\usepackage{float}
\usepackage[numbers]{natbib}
\usepackage{mdframed}
\usepackage{amssymb}
\usepackage{mdframed}
\usepackage[bookmarks=false]{hyperref}
\usepackage{multirow}
\usepackage{rotating}
\usepackage{enumitem} 
\usepackage{balance}

\usepackage{microtype}
\newcommand{\fref}[1]{Fig.~\ref{#1}}
\newcommand{\fig}[1]{Fig.~\ref{fig:#1}}

\newcommand{\tref}[1]{Table~\ref{#1}}
\newcommand{\tab}[1]{Table~\ref{tab:#1}}
\newcommand{\etal}{\textit{et al.}}

\newcommand{\bi}{\begin{itemize}}
\newcommand{\ei}{\end{itemize}}
\newcommand{\be}{\begin{enumerate}}
\newcommand{\ee}{\end{enumerate}}

\newcommand{\LH}{SHORT}

\usepackage[T1]{fontenc}
\usepackage{textcomp}
\usepackage{lmodern}

\usepackage[framed]{ntheorem}
\usepackage{framed}
\usepackage{tikz}
\usetikzlibrary{shadows}
\theoremclass{Lesson}
\theoremstyle{break}

\tikzstyle{thmbox} = [rectangle, rounded corners, draw=black,
  fill=gray!20,  drop shadow={fill=black, opacity=1}]

\newshadedtheorem{lesson}{Result}

\begin{document}

\title{``SHORT''er Reasoning About \\ Larger Requirements Models}
%\title{SHORT Reasoning About Large RE Models}
  \author{\IEEEauthorblockN{George Mathew, Tim Menzies}
    \IEEEauthorblockA{Computer Science,  NC State University, USA\\
      george.meg91,tim.menzies@gmail.com}
        \and
\IEEEauthorblockN{Neil A. Ernst, John Klein}
\IEEEauthorblockA{Software Engineering Institute, Pittsburgh, USA\\
nernst,jklein@sei.cmu.edu}}

 \maketitle
 \begin{abstract}
 %Requirements models support communication and decision-making. However, when requirements models get too complex, it becomes difficult for stakeholders to understand all their nuances.
 %As with humans,so too with  algorithms.
When Requirements Engineering(RE) models are unreasonably complex, they cannot  support efficient decision making.
SHORT is a tool to    simplify that reasoning by exploiting the  ``key''  decisions within RE models. These ``keys'' have  the property that once values are assigned to them, it is  very fast to reason over the remaining decisions.
Using these ``keys'',
reasoning about RE models can be greatly SHORTened  by focusing stakeholder
discussion on just these key decisions.

% One solution to this problem comes from AI where researchers report that many models have ``keys''. These are . %This paper shows that such keys exist in RE models, and can be used to simplify model-based RE. We explore this using a novel toolkit called SHORT.
This paper evaluates the SHORT tool on  eight complex RE models.
We find that the number of keys are typically only 12\% of all decisions. Since they are so few in number,
  keys can be used to reason faster about   models. For example,
using  keys,  we can  optimize over those models (to achieve the most goals at least cost)
two to three orders of magnitude faster than  standard methods.  %Significantly, the quality of the solutions found by SHORT are competitive with state-of-the-art optimizers.
Better yet, finding those keys is not difficult: SHORT runs in low order polynomial time and terminates in a few minutes for the largest models.
 
\end{abstract}


\pagestyle{plain}
\section{Introduction}
\label{sec:intro}
\begin{figure}[!b] 
  ~~~\includegraphics[width=3.1in]{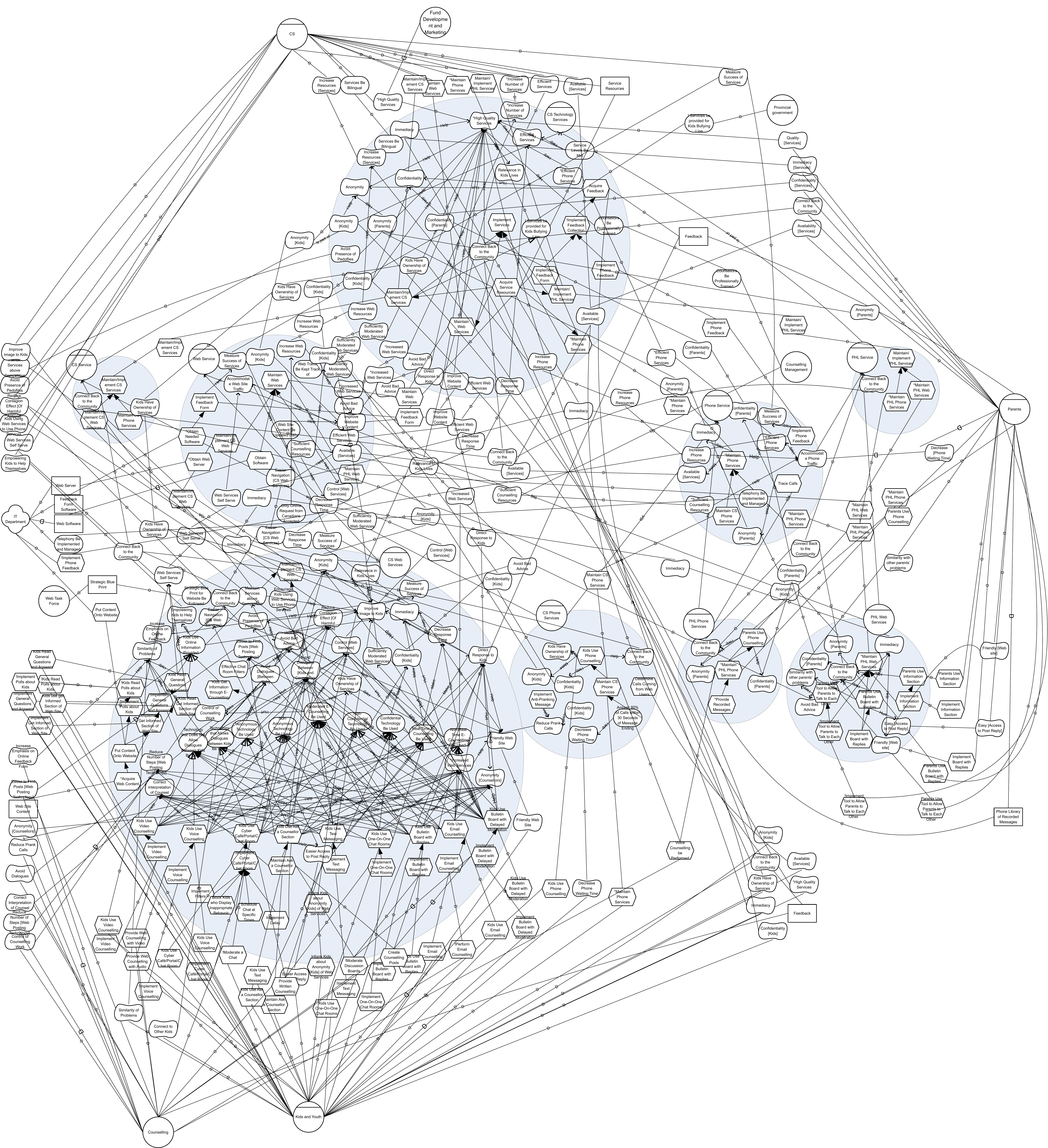} 
    \caption{Options for services in a CS department (i* format).  If the reader finds  this model confusing, then our point is made.  }
    \label{fig:cs}
\end{figure}

%
%When reasoning about complex requirements,   models are built to 
%documents  stakeholders beliefs.
Many researchers~\cite{mylopoulos92.nfr,Lamsweerde2001,amyot10,kaiya02,technere2010,yu97a,simon96,moran96} report that
the process of building and analyzing requirements  models   help stakeholders better understand the ramifications of their decisions.
However, complex models can  overwhelm stakeholders.
Consider a committee reviewing the goal model shown in \fig{cs}, that describes the information needs of a university computer science department~\cite{Horkoff2016}. This committee may have trouble with manually reasoning about all the conflicting relationships in this model. Further, automatic methods for the same task are hard to scale: as discussed below, reasoning about inference over these models is an NP-hard task.

But are models like \fig{cs} as complex as they appear? That graph is somewhat of a tangle---if we straightened out all the dependencies, would we find that the model depends on just a few {\em key} decisions? 
As shown below,  
such keys have been seen in other domains~\cite{williams2003backdoors,Amarel1986,Crawford:1994,Kohavi:1997}. 
This paper argues that it is beneficial to look for these keys. If they exist,
we can achieve ``shorter'' reasoning about large RE models, where ``shorter'' means
(a) large models can be processed in a very short time; (b) stakeholders can get faster feedback from their models;
(c)~stakeholders can conduct shorter debates about their models since they only need to debate a few key    decisions.

The rest of this paper hunts for keys in RE models.
Our next section offers a formal definition of keys,
plus a simple introductory example 
where  
 dozens of decisions are controlled by just  three keys. Next,
 a review of the   AI literature will show that
 that many models are known to have keys.
After that, we introduce SHORT,  a novel toolkit for finding and using keys.
SHORT is then tested on eight RE models from the
  i*~community~\cite{yu97a} (see \href{http://goo.gl/K7N6PE}{goo.gl/K7N6PE}).

 \begin{figure*}[!t]
%\begin{mdframed}
\begin{center}
    \includegraphics[width=6.6in]{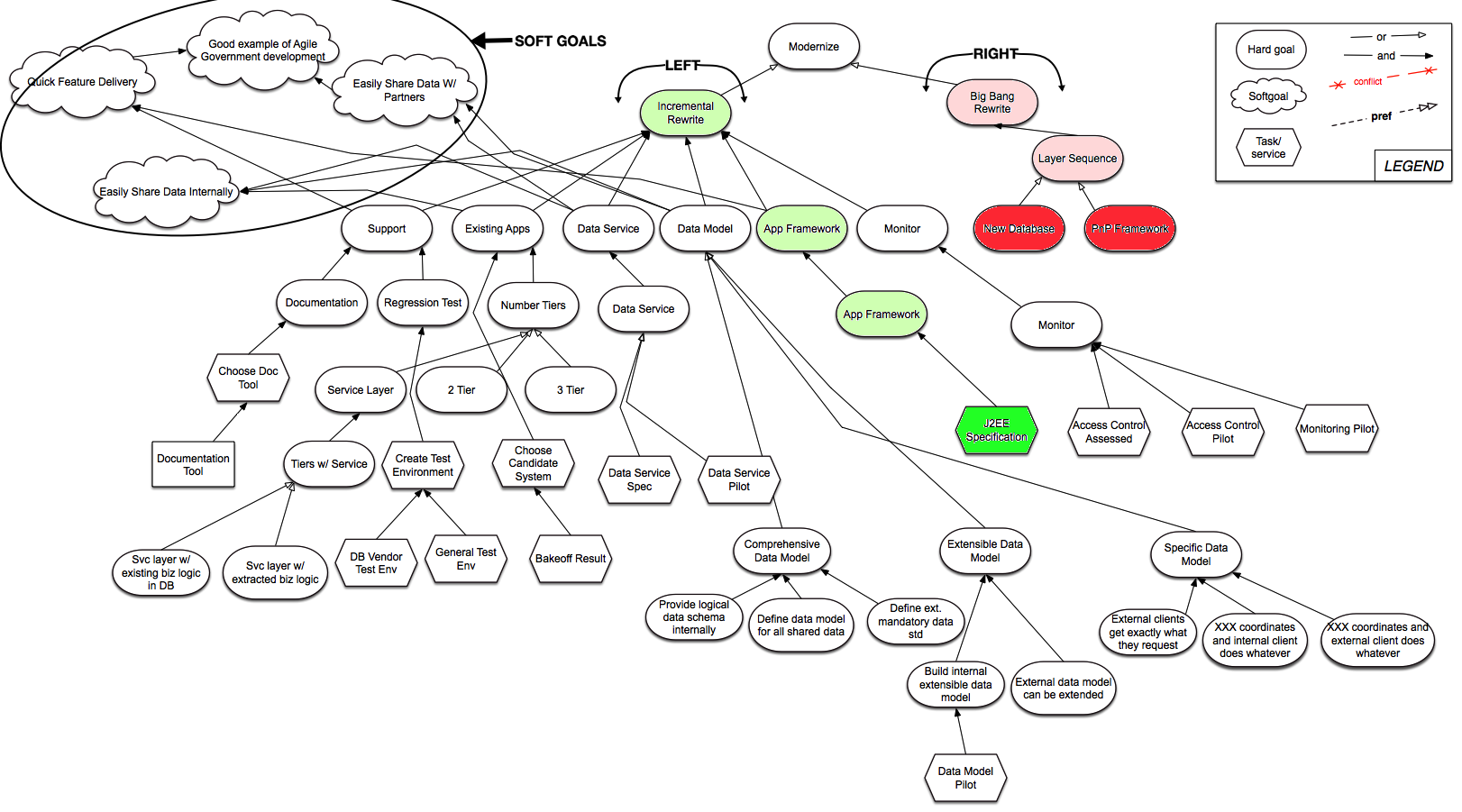}
\end{center}

%\end{mdframed}
\caption{A requirements model for
   IT System modernization.   This model has the syntax of \fig{sg}.  The three keys
of this model are the leaf nodes shown in red and green (see text for an explanation of why these are 
the keys). }
    \label{fig:aows}
\end{figure*}

  From those experiments, we conclude that 
 {\bf keys can be found in numerous
 RE models} since, in the sample of models explored here,
  the number of keys per model is   small: typically 12\% of all decisions.  
Also,  {\bf  keys are easy to find}. 
 SHORT   runs in low-order polynomial time
and terminates in just a few minutes (even for our largest model). This is a significant result:  all  prior attempts  suffered from crippling runtimes that prevented scale up~\cite{menzies1996applications,ernst12caise}.
Further,  {\bf it is very useful   to apply   keys for reasoning about RE models}.
 Using  the keys,  SHORT  can  find
decisions that satisfy the most goals at the least cost, 100s to 1000s  times faster than standard methods. Further, the decisions found by SHORT are competitive with (and sometimes even better
than) those found by standard optimizers.

\section{``Keys'': Definition and Example}
\label{sec:introeg}

    Consider a model  (e.g. \fig{cs} or \fig{aows}) where    objectives $O \in \{x,y,.....\}$ are determined by stakeholders' decisions $D$. After randomly selecting decisions, we would see  variance on the objectives $V_O$. We say that a small set of decisions $K \subseteq D$ ($|K| \ll |D|$) are  {\em key},  if after fixing  decisions $K$ and randomly assigning the rest of the decisions in $D - K$, the variance $V'_o$ drops significantly, i.e., $V'_O \ll V_O$.
        
        Just to say the  obvious: it is  useful to explore the keys first since, once the keys are set,
        there is  more certainty about the impact of the remaining decisions. Also, for models with discrete, not continuous
        output, entropy could be swapped for variance in the above definition.

For example,  \fig{aows} shows a  requirements model about IT System modernization tasks such as 
the Y2K problem (moving from 2 digit to 4 digit years);  reacting to vendor decisions to end-of-life operating systems and database products; and
 improving architectural support for new capabilities (e.g.,
 support for mobile devices).
The model in \fig{aows} comments on the refactoring, re-architecting, and  redesign  of  existing  systems.  The model has the syntax of \fig{sg}.  Note that the model has dozens of decisions and   a few top-level goals shown circled top right: ``Good Example ...'', ``Easily Share Data Internally'', ``Modernize''.  All of these nodes have a cost determined by stakeholders.
 If
the stakeholders are not sure of the exact costs, we allow them to define a range, and then conduct Monte Carlo simulations that sample
randomly across that range.
That said, for the rest of this paper we assume that cost of all the decisions is sampled from a triangular distribution (in our modeling language, this is a simple facet to change).

The two red and one green decisions in \fig{aows} are  \emph{keys}. To see this, note that
the right-hand-side of the model has no connection to the top goals. Hence,  two sensible
  decisions for this model would be to deny the leaves marked in dark red, thus disabling the right tree.
  As to the left-hand-side of the model, the leaf marked in dark green (``J2EE specification'') selects
  the shortest left-hand-side sub-branch with the fewest leaves. If the goal is to achieve the most high-level goals
  (at least cost), then it is also sensible to take this single dark green leaf. Why? Well,   making any other decision
  on the    left-hand-side of the model  will significantly add to the overall cost of the solution, since all those other decisions
  require multiple conjunctions of decisions. It should be noted that such a visual inspection can never be performed on larger models(like \fig{cs}) with hundreds of nodes and numerous cross tree constraints.

It is reasonable to ask what to do if users reject this analysis  and  explore the deeper branches on the left hand side. In such a case, we would  tell a requirements engineer that ``minimizing cost" is not a primary goal of these
users.  SHORT could then be rerun,  removing the ``minimizing cost" goal. That would lead to new alternatives, to  be debated by the stakeholders. Note that it would not be burdensome  since, as shown later(in Section \ref{sec:keysRE}), SHORT runs in just a few seconds.

 \begin{figure}\small
        \begin{tabular}{|p{.95\linewidth}|}\hline
        
      1.  {\bf Nodes:} have
    labels and  \emph{decisions}  are label assignments.
    
    2. {\bf Edges:}
          {\em Con} nodes connect to 
    other nodes via one four edges types ``makes, helps, hurts, breaks'' with
    weights \mbox{$E_j \in \{1,$\textonehalf$,-$\textonehalf$,-1\}$},
    respectively.

       3. {\bf Node types:}
         Nodes in a goal model can be of type {\em leaf}, {\em
      com}bine, or {\em con}tribute ({\em com} nodes have
   sub-types {\em and, or}).
 {\em Leaf} nodes are different to the
   rest since they have no children.
 On the other hand, when dealing
    with {\em and,or} nodes, it is expected to meet {\em all,one}
    (respectively) of the requirements in the child nodes.
 {\em Con}
    nodes divide into {\em softgoals}, which users are willing to
    surrender if need be, and {\em hardgoals} which users are more
    eager to achieve.\\

    4. {\bf Labels:}
 Nodes $N_i$ have labels
    $\{1,$\textonehalf$,0,-$\textonehalf$,-1\}$ for {\em satisfied}, {\em partially satisfied}, {\em undefined}, {\em denied} and {\em partially denied}.

         5.  Initially,   labels are {\em undefined} and are then
    relabelled {\em satisfied} or {\em denied} by some labeling
    procedure (e.g. see \fig{step}) during which labels may
    temporarily labeled {\em partially satisfied/denied}.\\

     \\\hline
     \end{tabular}
     \caption{Syntax of goal models.}\label{fig:sg}
    \end{figure}

\section{Related Work}
 
   \subsection{Complexity of Processing RE Models}\label{sec:compsim}

   Given multiple stakeholders writing assertions into a requirements model,
   it is likely that those stakeholders will generate models that can contain contradictions;
   i.e. incompatible assignments of labels to variables. For example, in \fig{aows},
   it would be a contradiction to assign {\em satisfied} and {\em denied} to the same variable.
   
   In traditional logic, if some set of assertions generates a
    contradiction then all the assertions are discarded as
    inconsistent.
   But in requirements models, when inconsistencies are detected,
    it is standard practice to focus on zones of agreement and
    avoid the parts of the model leading to the inconsistencies.
    Nuseibeh lists several strategies for this approach~\cite{nuseibeh1996and}
    including   {\em ignoring} (skip over edges that lead  to  contradictions);
 {\em circumventing} (``slip around'' inconsistencies;
    i.e.  if inference is blocked due to inconsistency, the inference can
     explore other avenues); and 
   {\em ameliorating}  (when conflicts cannot be avoided,
     it is prudent to try reduce the total number of conflicts).
    The problem with these tactics is that they can be very slow.
    Formally, the study of models with contradictions is the called
    ``abductive reasoning''. Poole's THEORIST
    system~\cite{Poole1994WhoCT} offers a clean logical framework of such reasoning.
    In that framework, a goal graph is a  {\em theory} that
     contains a small number of upper-most {\em goals}.
     When we reason about that theory, we make {\em assumptions}
     about either (a)~initial
     facts or (b)~how to resolve contradictory decisions.
     In the general case, only some subset of the theory can
     be used to achieve some of the goals using some of the assumptions without
     leading to contradictions (denoted $\bot$). That is:

     {\small \begin{equation}\label{eq:subs}
     \begin{array}{r@{~}c@{~}l}
       T &\subseteq& \mathit{theory} \\
       A &\subseteq& \mathit{assumptions}\\    
       G &\subseteq& \mathit{goals}\\
       T \wedge A&  \vdash & G\\
       T \wedge A & \not\vdash& \bot
       \end{array}
     \end{equation}}
     A {\em world of belief} is a solution that
     satisfies these invariants.
     For many years we have tried to find such worlds using a variety of
     methods. For example Menzies' HT4 system~\cite{menzies1996applications}
     combined forward and backward chaining to generate worlds while Ernst \etal~\cite{ernst12caise} used DeKleer's ATMS (assumption-based truth-maintenance system)~\cite{de1986assumption}.
Those implementations suffered
     from cripplingly slow runtimes that scaled very poorly to larger models. Such
     slow runtimes are not merely a quirk of those implementations---rather
     they are fundamental to the process of exploring models with many contradictions.
     It is easy to see why:
     exploring all the the subsets in Equation~\ref{eq:subs} is a very slow process.
     Bylander \etal~\cite{Bylander1991}  and
     Abdelbar \etal~\cite{Abdelbar:2004} both confirm that abduction is NP-hard;
     i.e. when exploring all options
within a requirements goal model, we should expect
very slow runtimes.

\subsection{Reports of ``Keys'' in AI Research}
\label{sec:keysAI}
     Just because an inference task is NP-hard, that does not
     necessarily imply that task will always exhibit exponential runtimes.
     Numerous AI researchers studying NP-hard tasks report the existence
     of a small number of {\em key} variables that determine the behavior of the rest
     of the model. When such keys are present, then the problem of controlling
     an entire model simplifies to just the problem of controlling the keys.

     Keys have been discovered
     in AI many times and called many different names:
     {\em Variable subset selection,
narrows,
master
variables}, and {\em backdoors}.
In the 1960s, Amarel observed that search problems contain
{\em narrows}; i.e. tiny sets of variable settings
that must be used in any solution~\cite{Amarel1986}.
Amarel's
work defined macros that encode paths between the narrows in
the search space, effectively permitting a search engine to leap quickly
from one narrow to another.

 \begin{figure*}
      \centering
      \includegraphics[width=\textwidth-4pt]{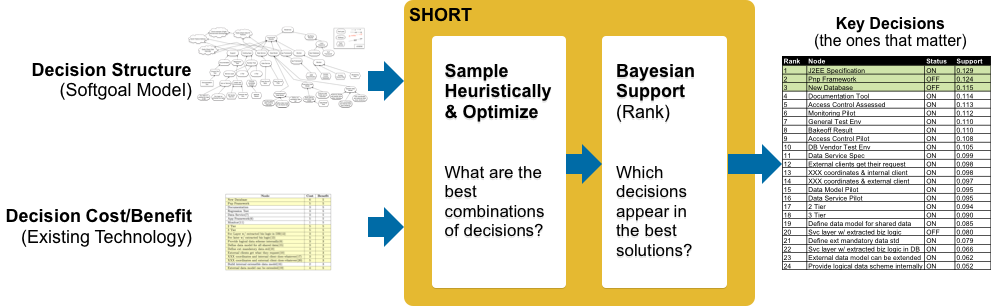} 
      \caption{Finding solutions with SHORT. Costs and benefits are inputs are domain-specific
      constructs that, in our work, we develop with users as part the model elicitation process.}
      \label{fig:workflow}
    \end{figure*}

In later work,  data mining researchers in the 1990s explored and examined what happens when a data
miner deliberately ignores some of the variables in the
training data. Kohavi and John report trials of data sets where  up to 80\% of
the variables can be ignored without degrading classification accuracy~\cite{Kohavi:1997}.
Note the similarity with Amarel's work: it is more important to reason about a small
set of important variables than about all the variables.

At the same time, researchers in constraint satisfaction found  ``random
search with retries'' was a very effective strategy.
Crawford
and Baker reported that such  searches
took less
time than a complete search to find more solutions using
just a small number of retries~\cite{Crawford:1994}.
Their ISAMP ``iterative sampler'' 
 makes random choices within a model until it gets ``stuck''; i.e. until further choices
do not satisfy expectations.
When ``stuck'', ISAMP does not waste time fiddling with current choices (as was done by older
chronological backtracking algorithms). Instead, ISAMP logs what decisions were made
before getting ``stuck''. It then performs a ``retry''; i.e. resets and starts again, this time making other random
choices to explore.

Crawford and Baker explain the success of this strange approach by assuming models contain
a small set of {\em master variables} that set the remaining variables (and this paper calls such master variables \emph{keys}).
Rigorously searching through all variable settings is not
recommended when master variables are present, since only a small number
of those settings actually matter. Further, when the master variables are spread thinly over the entire model,
it makes no sense to carefully explore all parts of the model since much time will be wasted ``walking'' between the far-flung master variables.
For such models, if the reasoning gets stuck in one region, then the best thing to do is to leap at random to some distant part of the model.

A similar conclusion comes from the work of
Williams \etal~\cite{williams2003backdoors}.
They found that if a randomized search is repeated many times,
that a small number of variable settings
were shared by all solutions. They also found that if they set those variables
before conducting the rest of the search, then formerly exponential runtimes
collapsed to low-order polynomial time. They called these shared variables the {\em backdoor}
to reducing computational complexity.
   
Combining the above, we propose
the following strategy for faster reasoning about RE models.
First,
use random search with retries to find the ``key'' decisions in RE models.
    Second,
    have stakeholders debate, and then decide, about the keys
     before exploring anything else.
Third,
to avoid trivially small solutions, our random search should strive
to cover  much of the model.

The rest of this paper implements this  strategy in a tool called SHORT.
This tool is
a multi-objective optimizer that seeks to maximize goal satisfaction, while
at the same time maximizing the softgoal satisfaction and minimizing the sum
of the costs of the decisions made within that model.

    \section{Goal Inference with SHORT}

 As summarized in \fig{workflow}, SHORT   runs in five phases:
    \bi
  \item[\textbf{SH}.] \underline{S}ample \underline{H}euristically the possible labellings.
  \item[\textbf{O}.] \underline{O}ptimize the label assignments in order to cover more goals
    or reduce the sum of the cost of the decisions in the model.
    To implement the  {\em retry} step recommended
     by Crawford and Baker's ISAMP \cite{Crawford:1994}. 
    \item[\textbf{R}.] \underline{R}ank all decisions according to how well they performed during the optimization process.
\item[\textbf{T}.] \underline{T}est how much conclusions are determined by the decisions that occur very early in that ranking.
    \ei

      {\bf SH = \underline{S}ample Heuristically:}
    The SAMPLE procedure of \fig{sample} finds one possible
    set of consistent labels within a goal.
    The procedure
    calls the STEP procedure (\fig{step}) over all the decisions in the model, labeling as many nodes as possible.
       This  procedure executes in a  random order
   since (a)~this
   emulates the idiosyncrasies of human discussions; (b)~as mentioned above,
   Crawford and Baker found this to be a useful strategy for reducing computational
   complexity.

Why use SAMPLE when there are many other soft goal inference methods
in the literature, such as e.g., 
   the forwards and backwards analysis proposed by Horkoff and Yu~\cite{Horkoff2016}? One reason we prefer SAMPLE is its generality. SAMPLE was first developed for Menzies' HT4
   system and has been applied to many    model types including
   a)~causal diagrams; (b)~qualitative equations; a (c)~frame-based knowledge representation;
               (d)~compartmental models, and even (e)~first-order systems (with finite-domains on the
               variables)~\cite{Menzies1996}.

     \begin{figure}\small
        \begin{tabular}{|p{.95\linewidth}|}\hline
    1. When considering an edge $E_j$ from $N_i$ to a child $N_k$
    then:
    \bi
 \item If $N_i\in \{1,-1\}$ then expect $N_k=E_j*N_i$  else
 \item If $E_j > 0$ then expect $N_k = N_i$  else
 \item If $E_j < 0 $    then expect $N_k = - N_i$.
   \ei
   Goal model inference does not use a fuzzy set or probabilistic approach to reasoning
   about conflicting influences. Rather, goal edges are either used or
   {\em ignored}. Hence:

   2. If $N_k$ is {\em undefined}, then it is labelled with the above expectations
   and we call STEP recursively
   over all edges to $N_k$'s children. Children are explored in a random order.
After recursion, for and-nodes, labels get set back to {\em undefined} if  any fail  $N_k$  expectations (as defined by point \#1).

   3. Otherwise, if its  label does not meet the above expectations, then
   we {\em ignore} the edge $E_j$.
   \\\hline
     \end{tabular}
     \caption{Procedure STEP: labels neighboring nodes; may {\em ignore}
       some edges. Called by the SAMPLE procedure of \fig{sample}.}
  \label{fig:step}
\end{figure}

 \begin{figure}\small
     \begin{tabular}{|p{.95\linewidth}|}\hline
     1. SAMPLE inputs (a)~a goal model and (b)~a set of {\em prior}
     decisions made by any previous run of SAMPLE (so initially,
     this set is empty).

   2. As it initializes, SAMPLE sets all nodes to {\em undefined} then
   all {\em prior} nodes  to {\em satisfied};

   3. While there are {\em undefined} nodes,
   SAMPLE (1)~``picks'' one  {\em  undefined} decision as {\em satisfied}
   then    (2)~``reflects'' over its edges using the STEP procedure
   of~\fig{step}.
   SAMPLE is stochastic since  ``picks'' and ``reflects'' return nodes and edges
   in a random order.

   4. SAMPLE outputs a {\em solution}  listing all   {\em satisfied nodes}.
   \\\hline
   Important note:
   if {\em prior} is changed, then SAMPLE will return different solutions.
   That is, the results from SAMPLE can be carefully
   tuned and improved by the OPTIMIZE procedure of \fig{optimize}
   that carefully selects useful
   {\em prior}s.
     \\\hline
     \end{tabular}
     \caption{Procedure  SAMPLE: tries to labels many nodes.
     Called by the OPTIMIZE procedure of \fig{optimize}.}\label{fig:sample}
     \end{figure}
   \begin{figure}[!t]\small
     \begin{tabular}{|p{.95\linewidth}|}\hline
       1. Given a model with $n$ decisions,
       OPTIMIZE calls SAMPLE $N=10*n$ times.
       Each call generates one member of the population {\em pop$_{i\in N}$}.

       2. OPTIMIZE scores each {\em pop}$_i$ according to various objective
       scores $o$. In the case of our goal models, the objectives are $o_1$ the sum of the cost
     of its decisions, $o_2$ the number of ignore edges, and the number of $o_3$ satisfied goals
     and $o_4$  softgoals.

     3. OPTIMIZE tries to each replace {\em pop}$_i$ with a mutant $q$
     built by extrapolating between three other members of population $a,b,c$.
     At probability $p_1$, for each decision $a_k \in a$, then
     $m_k= a_k \vee (p_1 < \mathit{rand}() \wedge( b_k \vee c_k))$.

     4. Each mutant $m$ is assessed by calling  $\text{SAMPLE}(\textit{model,prior=m})$;
     i.e. by seeing what can be achieved within a goal after first assuming
     that $\textit{prior}=m$.

     5.  To test if the mutant $m$ is preferred to {\em pop}$_i$, OPTIMIZE uses
      Zitler's continuous domination {\em cdom}
      predicate~\cite{Zitzler2004}. This predicate compares two sets of objectives
      from sets $x$ and $y$. In that comparison,
      $x$ is better than another $y$ if $x$  ``losses'' least.
      In the following, $``n''$ is the number of objectives and $w_j \in \{-1, 1\}$
shows if we seek to maximize $o_j$.
\[
\begin{array}{rcl}
x \succ y & =& \textit{loss}(y,x) > \textit{loss}(x,y)\\
\textit{loss}(x,y)& = &\sum_j^n -e^{\Delta(j,x,y,n)}/n\\
\Delta(j,x,y,n) & = & w_j(o_{j,x}  - o_{j,y})/n
\end{array}
\]

5. OPTIMIZE repeatedly loops over the population, trying to replace  items with mutants,
until new better mutants stop being found.
\\\hline
\end{tabular}
     \caption{Procedure OPTIMIZE: strives to find ``good'' priors which,
       when passes to SAMPLE, maximize the number of edges used
       while also minimizing cost, and
  maximizing satisfied hard goals and soft goals.
  OPTIMIZE is based on Storn's differential evolution optimizer~\protect\cite{storn1997differential}.
  OPTIMIZE is called by the RANK procedure of \fig{rank}.
  For the reader unfamiliar with the mutation technique of step 3 and the {\em cdom}
  scoring of step~5, we note that these
  are standard practice in the search-based
  SE community~\cite{Fu2016,krall2015gale}.
}\label{fig:optimize}
\end{figure}

{\bf O = \underline{O}ptimize:}
\fig{sample} discusses how we SAMPLE one set of labels from a softgoal model.
As described at the bottom of that figure, that SAMPLE-ing process can
be controlled via the {\em prior} decisions passed to the model.
 The OPTIMIZE procedure of \fig{optimize} is a multi-objective evolutionary algorithm that
  learns which {\em
   prior}s  decisions lead to {\em better} labellings. Here, a {\em better} labelling produces (a)~greater coverage of goals and softgoals,
     (b)~minimization of skipped edges; (c)~least cost solutions.  Note: for
     this paper, we sample our decision costs from a triangular distribution (in future
     work, we will explore other  distributions).

   {\bf R= \underline{R}ank:}
         The RANK procedure of \fig{rank}
         reflects of the results of the optimizer to rank
         each decisions. Decisions are ranked by the probability that they are
         associated with the better goals.
         Note that the key decisions will be 
         among the highest ranked decisions.

         {\bf T = \underline{T}est:}
         The TEST procedure of \fig{test}
         takes a ranked list of decisions,
         then tests what happens when the first few  are
         fixed and the rest are selected at random.

              \begin{figure}\small
       \begin{tabular}{|p{.95\linewidth}|}\hline
         1. Run OPTIMIZE $N=20$ times, keep all decisions $d$ in the generated population {\em pop$_{i \in N}$}
         and their  objectives $o$.

         2. For each objective $o_j \in o$, sort and separate  the top 10\% $o_j$ scores; call that ``best'' $b$ and the
         remainder ``rest''.
  For each  decision $d_k \in d$, do
          \bi
          \item Count how many times $n_1$ that $d_k$ is associated with a ``best'' objective
            score. Set $n_2=N-n_1$
          \item The value of $d_k$ for objective $o_j$ is the probability $p$ times the support $s$
              that $d_k$ appears more often in ``best'' than ``rest''.
              That is, $s_{j,k}=n_1*0.1$ and  \mbox{$p_{j,k} = (n_1*0.1)/(n_1*0.1+n_2*0.9)$}.
            \ei
    
              3. Let {\em ordering} be  all decisions, sorted descending by
              the value of decision $d_k$ across all objectives; i.e.
              $v_k = \sum_j s_{j,k} \times p_{j,k}$\\\hline
       \end{tabular}
       \caption{Procedure RANK:  ranks all decisions according to how well they performed during the optimization process. Used by the TEST procedure of \fig{test}.}
       \label{fig:rank}
       \end{figure}

     \begin{figure}[!t]
       \small
           \begin{tabular}{|p{.95\linewidth}|}\hline
             1. Run RANK to get a sorted list of decisions $d$.
              2. For all decisions $1 \le x \le |d|$, do
              \bi
              \item Build a {\em prior} set using decisions $d_1 .. d_x$ from {\em ordering};
              \item 20 times, call  {\em SAMPLE(model, prior)}
              \item Record for position $x$ the median and IQR of the objectives found in those results.
                \ei
                Aside: median= 50th percentile; IQR=(75th-25th) percentile.\\\hline
          \end{tabular}
           \caption{Procedure TEST: from a ranked
             list of decisions $d$, set the  first few decisions,
             then   selected the rest
             at random.}
           \label{fig:test}
     \end{figure}

  \begin{figure}
    \centering
     \includegraphics[width=3.5in]{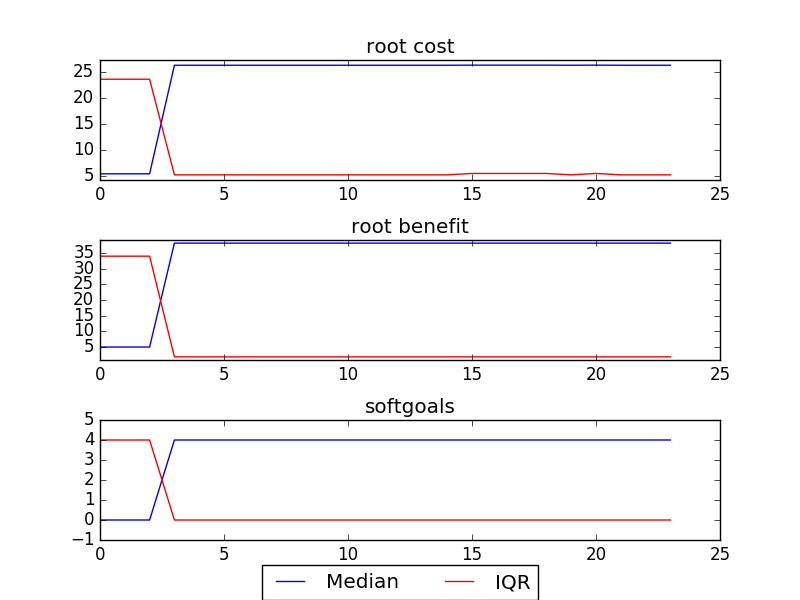}
    \caption{
      The X-axis contains
      $d$ decisions sorted by the RANK procedure of \fig{rank}.
      The y-axis shows the results from the TEST procedure of \fig{test};
      i.e. for $1 \le x \le |d|$, fix the first $x$ decisions
      then, 20 times, make random choices about
      decisions $x+1$ to $|d|$.
      Results shown as median and IQR values.
      Median = 50\% percentile and IQR=intra-quartile range= (75-25)th percentile.
  These results are  {\em smoothed} such that these plots only change
  where this is statistically significant non-small change in the y-axis.
  An appendix to this paper describes our statistical smoothing procedure.
      }\label{fig:neil1_star1}
\end{figure}

   \subsection{Reporting the Results}
   The SHORT process described above is a large-scale
   ``what-if''+optimization procedure. In order to succinctly
   describe the results of the above, we apply the following
   statistical summarization method.

    \fig{neil1_star1}  shows what is reported by TEST after analyzing the modernization
    model of \fig{aows}.
    At each point along the x-axis, TEST  samples
the goal model 20
    times using {\em prior} decisions taken from $1 \le i \le x$.  In the
    solutions returned by SAMPLE, {\em root cost} is the sum of decision costs;
    {\em root benefit/softgoals} are the sum of the number of satisfied
    goals/soft goals. The blue and red plots show the median and variation around
    the median.

     In these results,   after
  making three decisions, the medians rise to a steady plateau and
  the variations plummet. That is,  the model in \fig{neil1_star1} contains keys, as defined earlier in Section \ref{sec:introeg}.

This figure is reporting that this model
 has three key decisions which, if set, make further discussion
 superfluous.  Note that these results match what we learned from a visual inspection of \fig{aows} in Section \ref{sec:introeg}.

 \begin{figure}[!t]
    \begin{tabular}{|p{.95\linewidth}|}\hline
    \scriptsize
         \be
       \item J2EE Specification     (satisfied)
         \item \emph{Pnp Framework     (denied)}
\item  \emph{New Database     (denied)}
\item  Documentation Tool (satisfied)
\item  Access Control Assessed (satisfied)
\item  Monitoring Pilot     (satisfied)
\item  General Test Env     (satisfied)
\item  Bakeoff Result     (satisfied)
\item  Access Control Pilot (satisfied)
\item  DB Vendor Test Env (satisfied)
\item  Data Service Spec     (satisfied)
  \item External clients get their request (satisfied)
\item  Co-ordinates \& internal client (satisfied)
\item Co-ordinates \& external client (satisfied)
\item Data Model Pilot     (satisfied)
\item Data Service Pilot     (satisfied)
\item  2 Tier     (satisfied)
\item  3 Tier     (satisfied)
\item Define data model for shared data (satisfied)
\item  \emph{Svc layer w/ extracted biz logic (denied)}
\item  Define ext mandatory data std (satisfied)
\item  Svc layer w/ extracted biz logic in DB (satisfied)
\item  External data model can be extended (satisfied)
\item  Provide logical data scheme internally (satisfied)
  \ee\\\hline
  \end{tabular}
  \caption{Results of decision ordering on \fig{aows}. Read ``satisfied/denied'' as equivalent to
  ensuring that a leaf node is achieved or not achieved.
     Note the {\em denied} items at positions $x\in
\{2,3,20\}$. These are recommendations of what {\em not} to do. The
rest are all positive recommendations. The implication of \fig{neil1_star1}
is that after following the recommendations for the first 3 decisions, the remaining
decisions will have little additional impact on overall cost or satisfaction.}
\label{fig:record}
\end{figure}

The decision rankings found by RANK for \fig{aows}
are shown in \fig{record}.
Two features of this list deserve  comment.
Firstly,
  this list is {\em not} 24 decisions that could
  inspire $2^{24}>16,000,000$ debates.  Rather, it is an {\em
    ordering} with the property that item $x+1$
  is recommended {\em only} if recommendations $1..x$ are first adopted.
  That is, this list offers only 24 decisions to users (if they choose to or not to perform actions $1..x$).
Secondly,
  if users cannot implement all these recommendations, they
  can easily read from \fig{neil1_star1}  the effects of implementing
  just the first $x$ items.

% \begin{figure}[!t]
% %\begin{mdframed}
% \begin{center}
%     \includegraphics[width=3.5in]{AOWS_modified}
% \end{center}
% \caption{Highlighting the  key decisions of \fig{aows}.
% As per the first three recommendations of \fig{record},
% nodes in dark red a {\em denied} and nodes in dark green are {\em satisfied}. All the other colors
% are decisions that follow on from these three key decisions.}\label{fig:why}
% \end{figure}

 \begin{table}[!t]
\footnotesize
\centering
\caption{Goal models used in this study, sorted by number of edges.}
\begin{tabular}{r|rrl}
 \textbf{Model} & \textbf{Nodes} & \textbf{Edges} &    \\ \hline
Services & 351 & 510 & (see \fref{fig:cs})  \\
Counselling & 350 & 470  & \\
Marketing & 326 & 422 & \\
Management & 206 & 239  & \\
ITDepartment & 126 & 162  & \\
Kids\&Youth & 81 & 81 &    \\
IT Modernization & 53 & 57 & (see \fref{fig:aows})\\
\end{tabular}
\label{tab:sizes}
\end{table}

\section{Validation: Looking for Keys in RE Models}
\label{sec:keysRE}
We applied SHORT to 8 real-world RE models as an exploratory validation of our belief in the presence of keys in other RE models. We then compared SHORT to NSGA-II~\cite{Deb02}, a state-of-the-art optimizer.

         The introductory model in \fig{aows} 
         has 53 nodes and 57 edges.
         Table~\ref{tab:sizes} shows some details
         on the other goal models used in this study.
          The largest of our sample
         is the {\em CSServices} models shown in \fig{cs}.
         For images of all these models, see \href{https://goo.gl/K7N6PE}{goo.gl/K7N6PE}.
        These models were used since they are the  largest publicly accessible RE models .
All these models conform to the syntax defined  in \fig{sg}.

\input{obj_reports.tex}

         \tab{partials} shows  SHORT's results
          after making
         the first
         6,12,25,50,100\% of the decisions along the ranking
         found by RANK.  In a result
         consistent with \fig{neil1_star1},  there is little change to goal or soft goal coverage
         are making just a few decisions:          typically, just  12\% of the decisions(see \href{https://goo.gl/cRv9Lm}{https://goo.gl/cRv9Lm}).

         Two exceptions to the pattern ``12\% is enough''
         are
         CSCounselling and ITDepartment. Those models achieved nearly max goal coverage
         at 12\% but did not peak until making 25\%         of the decisions.
         Even with that exception, the general conclusion is clear: with SHORT,
         only a minority of the decisions need to be made with care, since once those
         are made, the goal coverage is robust (unchanging) for the remaining decisions.

    (Aside: one quirk of \tab{partials} is that, as shown in the column
    headed with ``0'', even when no
    decisions are made, it is possible to achieve
    some of the goals and softgoals, just by making decisions at random.
    Since no human has committed to any decision
    in that region, we mark all the 0\% costs as ``n.a.'' to indicate that they are not applicable.)
    
\begin{figure}
    
    \includegraphics[width=0.95\linewidth]{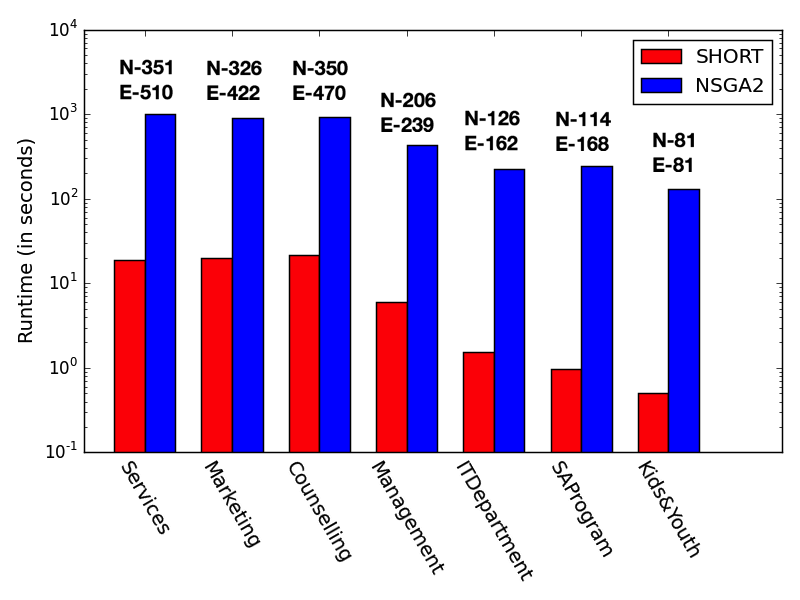}
    \caption{Runtimes for nsga2 compared to our approach. The number of nodes(N) and the edges(E) for each model is shown above the respective bar.}
    \label{fig:sg_runtimes}
\end{figure}

  \fig{sg_runtimes} shows the runtimes required to generate our results.
         Empirically, these runtimes fit the curve $\mathit{secs}=\mathit{nodes}^2)$
         with an $R^2=0.97$; i.e. SHORT's runtimes are low-order polynomial.
         This is a significant result  since in 1990, Bylander \etal
 warned that abductive search is NP-hard~\cite{Bylander1991};
i.e. when exploring all options
within an RE model, we should expect
very slow runtimes. The pessimistic
result has confirmed empirically by Menzies \etal ~\cite{Menzies1996305}, theoretically by Abdelbar \etal~\cite{Abdelbar:2004},
and then again empirically by
Ernst \etal~\cite{ernst12caise}.

It is insightful to compare SHORT's runtimes against the {\em forward}
and {\em backwards} analysis proposed by
Horkoff and Yu~\cite{Horkoff2016}.   They report times in the range of 7 to 300 seconds (which
includes human time considering various choices)---which is
approximately the same as the runtimes seen in \fig{sg_runtimes}.
The advantage of backward and forward analysis is that user involvement
increases user acceptance of the conclusions. The disadvantage is that
those analysis methods do not comment on the robustness of the solution.
Further, a forward and backwards analysis results in one sample of the model.
A SHORT-style analysis, on the other hand,
includes extensive ``what-if'' simulations. Charts like
\fig{neil1_star1} not only offer solutions but also comments
on:
\bi
\item
  The stability of the selected decisions: see the red ``IQR'' line
  in that figure;
\item
  The trade space across multiple decisions. Users can check a
  display of SHORT's decision orderings (like  \fig{neil1_star1})
  to decide for themselves when enough benefit has been obtained for enough cost.
  \ei
  To the best of our knowledge,
  forward and backward analysis
  has not been benchmarked against alternate techniques.
  
  Our next results compare SHORT against a state-of-the-art multi-objective optimizer called NSGA-II~\cite{Deb02}.
  This is a widely used genetic algorithm that uses a novel {\em select} operator to find the best ``parents''
  to make the next generation.
  For both NSGA-II and SHORT, the goal is to maximize the coverage of goals and soft goals, while at the same time
  minimizing the sum of the costs of the decisions.  \fig{sg_runtimes} compares the median runtime for {\em one} run of NSGA-II and SHORT. Note that
SHORT runs 2-3 orders of magnitude times faster.

In addition, NSGA-II  returns just one  solution, whereas SHORT  reports   what happens when an increasing number
  of decisions are imposed on a system (as done in \fig{neil1_star1}). In order for NSGA-II   to reason like SHORT, it
  would have to run hundreds of `what-if'' studies. That is,  NSGA-II's runtime costs would be incurred hundreds of times.
  
  % One reason to prefer SHORT is that, as shown in \fig{sg_runtimes}, SHORT completes its version of that ``what-if'' study in just a few minutes.
To explain SHORT's faster runtimes, we invoke the same reasoning
  used by Crawford and Baker, which we described in Section \ref{sec:compsim}.
The RE models contain a small set of key variables  that  determine the rest.
Hence,
rigorously searching through all variable settings is not
recommended   since only a small number
of those settings actually matter. Further, when the key decision
variables are spread thinly over the entire model,
it makes no sense to carefully explore many parts of the model (as done by NSGA-II) since much time will be wasted ``walking'' between the far-flung keys.
For such models, if the reasoning gets stuck in one region, then the best thing to do is to leap at random to some distant part of the model (as done by SHORT's random search).

  \tab{nsga2_small} compares how objectives were covered in 20 repeated runs of NSGA-II and SHORT.
  Both approaches achieved remarkably similar coverage
  of goals---an effect that can be explained by our models having a small number of key variables
  which, if set the right way, control what can be achieved in the rest of the model.

  \begin{table}\scriptsize
\centering
\begin{tabular}{|l|l|l|}
\hline
\textbf{Model} & \textbf{NSGA2} & \textbf{SHORT} \\ \hline
Services & \begin{tabular}[c]{@{}l@{}}f1: 46.77 $\pm$ 0.0\\ f2: 65.22 $\pm$ 0.0\end{tabular} &  \begin{tabular}[c]{@{}l@{}}f1: 46.24 $\pm$ 0.0\\ f2: 65.22 $\pm$ 0.0\end{tabular} \\ \hline
Marketing & \begin{tabular}[c]{@{}l@{}}f1: 32.47 $\pm$ 0.0\\ f2: 34.38 $\pm$ 0.0\end{tabular} & \begin{tabular}[c]{@{}l@{}}f1: 32.47 $\pm$ 0.0\\ f2: 34.38 $\pm$ 0.0\end{tabular} \\ \hline
Counselling & \begin{tabular}[c]{@{}l@{}}f1: 54.29 $\pm$ 1.41\\ f2: 44.83 $\pm$ 3.45\end{tabular} & \begin{tabular}[c]{@{}l@{}}f1: 53.59 $\pm$ 2.76\\ f2: 44.83 $\pm$ 3.45\end{tabular} \\ \hline
Management & \begin{tabular}[c]{@{}l@{}}f1: 44.63 $\pm$ 0.82\\ f2: 61.11 $\pm$ 2.78\end{tabular} & \begin{tabular}[c]{@{}l@{}}f1: 42.98 $\pm$ 1.65\\ f2: 61.11 $\pm$ 2.78\end{tabular} \\ \hline
ITDepartment & \begin{tabular}[c]{@{}l@{}}f1: 68.42 $\pm$ 2.63\\ f2: 69.57 $\pm$ 0.0\end{tabular} & \begin{tabular}[c]{@{}l@{}}f1: 68.42 $\pm$ 2.63\\ f2: 73.91 $\pm$ 8.69\end{tabular} \\ \hline
SAProgram & \begin{tabular}[c]{@{}l@{}}f1: 78.69 $\pm$ 1.64\\ f2: 66.67 $\pm$ 0.0\end{tabular} & \begin{tabular}[c]{@{}l@{}}f1: 78.69 $\pm$ 1.64\\ f2: 66.67 $\pm$ 0.0\end{tabular} \\ \hline
Kids\&Youth & \begin{tabular}[c]{@{}l@{}}f1: 22.86 $\pm$ 0.0\\ f2: 100.0 $\pm$ 0.0\end{tabular} & \begin{tabular}[c]{@{}l@{}}f1: 22.86 $\pm$ 0.0\\ f2: 100.0 $\pm$ 0.0\end{tabular} \\ \hline
\end{tabular}
\caption{NSGA-II compared to our SHORT approach. Columns report Accuracy(Median $\pm$ IQR) of the objectives \textit{f1:} Percentage of soft goals satisfied and \textit{f2}: Percentage of goals satisfied.}
\label{tab:nsga2_small}
  \end{table}

\section{Discussion}
                   
\subsection{Can SHORT be applied to other modeling languages?}
 
For our experiments, we used the models of  \tref{tab:sizes} for two reasons. First, they are the largest requirements models we can access.  
Second, these models are representative of a large class of other requirements languages; i.e. those that can be compiled into what we call PCBN; i.e. {\em \underline{P}ropositional assertions}, where subsets of the propositions are augmented with {\em \underline{C}ost}s, {\em \underline{B}enefit}s, and {\em \underline{N}ogood}s:
\bi
\item
    The use of {\em Cost}s was discussed, earlier in this paper in Section \ref{sec:introeg}.
\item
{\em Benefit}s relate to a system's high-level goals. For example, in \fig{aows}, the lower leaves have zero benefit while the goals circled top-left have substantial benefit.
\item    {\em Nogood}s report what variable assignments are not allowed. For example, if a variable has   mutually exclusive values, then it would be {\em nogood} for that variable to be given two different values.
\ei
Many modeling notations can be converted into PCBN. Menzies' HT4 system included a domain general {\em knowledge compiler} that generated PCBN from (a) causal diagrams; (b) qualitative equations; a (c)~frame-based knowledge representation; (d) compartmental models, and even (e) first-order systems (with finite-domains on the variables)~\cite{Menzies1996}.

Also, there are many recent examples of such knowledge compilation in the RE community. For example, any RE researcher using SAT solvers must write a knowledge compiler to convert their high-level notations into PCBN. Some of that research comes from within the i* community (e.g.,~\cite{wang2009self}) and some from elsewhere. For example, see researchers using SAT solvers to explore van Lamsweerde's goal graphs~\cite{van2008requirements}; requirements for software product lines and feature models~\cite{classen2008s,Mendonca:2009}, as well as other RE tools~\cite{benavides2007fama}. For a discussion on other RE models that can be compiled to propositions, see~\cite{technere2010}.

Last, we have some evidence that keys exist in non-PCBN models. The COCOMO tool suite is a system of numeric equations
that offer predictions for software development time, effect, defects and risks.  Menzies \etal ~\cite{menzies2007business} found
that these systems have ``keys'' as we defined in Section \ref{sec:introeg}.
In other work,   we have documented key-like effects in procedural systems~\cite{Menzies2012}.

\subsection{Why offer a new reasoning tool when others  exist? }

Many researchers use SAT solvers to find a solution within softgoals~\cite{sebastiani04caise}. For example,  Horkoff and Yu, in  \cite{Horkoff2013} discuss local propagation methods for reasoning backwards from goals or forwards from assertions across models like \fig{aows}  (and sometimes,  the Horkoff and Yu methods call SAT solvers as a sub-routine). Using SMT provers, these SAT solver methods can  handle 1000s of elements (see the recent work of  Nguyen \etal ~\cite{Nguyen2016}). 

Where SHORT differs from the above is that those methods find {\em one} solution while SHORT tries to build a trade space that summarizes the effects of decisions across  {\em all} solutions. As discussed in Section \ref{sec:compsim}, an {\em all} solution approach is NP-hard and, prior to this paper, all reported attempts to address this problem have suffered from crippling runtimes that that prevented scale up ~\cite{Amarel1986,Crawford:1994}. SHORT is the first implementation we have found (in the last 20 years) that achieves low-order polynomial runtimes for large RE models.

That said, there could be advantages to {\em combining} SHORT with a forward or backwards analysis.
Since SHORT is so fast, it could be run prior to a forward/backwards analysis in order to generate
the trade space.
Hence,
stakeholders could then use it as a guide while debating trade-offs during the forward/backwards analysis. 
This could be  one very promising future direction for this 
research.
 
% \subsection{ Are these ``keys'' just a quirk of the model in \fig{aows}?}  
 
%  As discussed in the section~\ref{sec:keysAI}, there is much evidence that suggests that the results from \fig{aows} will generalize to many more models. In the field of AI, many other research areas have reported that models often contain a small number of key decisions   (see discussion \tion{compsim}); and our own work has shown similar findings in  software-specific areas---see ~\cite{Menzies2012, menzies2004many,druzdel94}.
%  Further, as shown in the Section~\ref{sec:keysRE} at the end of this paper, these keys are common in RE models. 

% \subsection{How can SHORT be applied to Requirements Engineering in practice?}

\subsection{What other goals beside minimizing costs can be used as an input and how can you identify them?}

There are probably as many metrics as user goals in Requirements Engineering. For example, a) the functional and non-functional requirements~\cite{chung2012non} or b) the meta-requirements like cost, reused components, least defective, most different to existing system, etc. Methods for finding such engineering goals have been researched extensively in the qualitative software engineering literature, like summarizing opinions using card sort~\cite{yuan2015users} or the knowledge acquisition literature~\cite{guzzi2013communication}. SHORT is indifferent to the nature of the objectives the user aims to optimize, and can use any ordinal or continuous variables. 
% zaman2015penan, kropp2016interactive,bacchelli2013expectations, 

\subsection{Threats to Validity}
As with any empirical study, biases can affect the final
results. Therefore, any conclusions made from this work must
be considered with the following issues in mind:

\emph{Sampling bias} threatens any  experiment;
i.e. what matters in (say) practitioner settings may not be true of our examples.
The data sets used here comes from goal models taken from a repository
and all but one were supplied by one individual. These were all from i* style models.
That said, as argued above, the models we use here share many aspects
with RE representations used by many other researchers. The \fig{aows} model was derived from consulting work done at the SEI with a large US government agency, and is ongoing. See \cite{Ernst2016} for more background.

% \emph{Learner bias: }
% % For building the defect predictors in this
% % study, we elected to use k-Nearest Neighbor. We chose the kNearest
% % Neighbor because its results were comparable to the
% % more complicated algorithms [38] and can act as a baseline
% % for other algorithms.
% Classification is a large and active field
% and any single study can only use a small subset of the known
% classification algorithms. We chose NSGA-II since it is well-known search

\emph{Evaluation bias: }
This paper uses one measure of success, goal achievement and associated cost and benefit. Optimal solutions only consider inputs given, and may not reflect all the complexities of a given decision.

\emph{Construct Validity}: The conclusions of this paper are about   better processes for making
{\em decisions} (specifically, do not waste time on all the numerous redundant issues).
It would be a violation of construct
validity to make another claim---that
\LH{}-guided decisions lead to better outcomes.

% \subsection{Search-Based Software Engineering}
% The other related to this paper is the use of search-based techniques, such as our use of Differential Evolution, to find near-optimal solutions. Examples of this work
% include  the Next-Release Problem studied
% in the search-based software engineering community, for example, in Bagnall \etal \cite{Bagnall2001}. There, the problem is to determine the near-optimal set of requirements to prioritize in the next software release, given constraints such as cost and schedule. These formulations, however, tend to lack the intersectionality of the goal models we show above; dependencies rarely exist between requirements, and a simple cost metric is presumed to exist. Recently, work such as that by Tonella \etal \cite{Tonella2013173} has focused on amending the straight-forward search approach with expert input, to derive a more sophisticated prioritization. We build on this work by adding in the notion of model dependencies, but use their stakeholder prioritization in our earlier steps to derive initial values for the goals (not discussed in this paper).

% A possible extension here is to consider the model development as a series of steps in refining model uncertainty. Then the keys we find are areas of greater uncertainty, and more effort is needed to refine that portion of the model. This is similar to the notion of partial models expressed in Salay \etal \cite{salay12}.
% % NEil - find the old paper that actually discused this e.g. italian work and next release problem.
% % XXX add \cite{gay2010finding}?

\subsection{Future Work}
\label{sect:future}
SHORT can be expanded to work with quantitative goal modeling frameworks like KAOS~\cite{van2009requirements}. 

The largest goal models used in this study has 351 nodes and 510 edges (refer to \fref{fig:cs} and \tref{tab:sizes}), and from \fref{fig:sg_runtimes} we can see that this model can be reasoned in around 30 seconds compared to a classic optimization technique like NSGA2 (which takes around 900 seconds). Synthetic requirements engineering models can be generated by simulation~\cite{seybold2004evolution} and can be used to demonstrate a greater scalability of SHORT.

SHORT provides a framework for visualizing the ``keys'' for an RE model but it requires the model to be encoded using the i* notation. This can be alleviated by defining a a uniform encoding scheme to support both qualitative and quantitative models. This scheme can then be used to power a Graphical User Interface  tool to better aid stakeholders to develop models.

\section{Conclusions}

Confusing models can confuse stakeholders. One way to untangle complex  models is to find the small number of key decisions that  determine what can be done for the rest of the model.

Such keys are naturally occurring in   RE models. For example, in eight large RE models sampled from the the i* community, we have found  a small number of decisions (often, just 12\%)  was enough to to set the rest of the decisions. In stark contrast to much prior work~\cite{menzies1996applications,ernst12caise}, finding these keys was very fast, using the randomized search methods employed in SHORT. Using these keys, SHORT is able to optimize RE models in time that was orders or magnitude faster than state-of-the-art optimizers. Further, the optimizations found by SHORT were competitive with those found by other, much slower, methods. Hence, our conclusions
are that:
\be
\item  Keys can be found in numerous RE models;
\item Keys are easy to find;
\item It is very useful to apply keys for reasoning about RE models. 
\ee
% As mentioned in the the introduction, these
%   results do {\em not} show that  all requirements models simplify to small
% number of key decisions. However:
% \bi
% \item
%   When keys occur in RE models, then
%   model-based RE can be greatly simplified by focusing stakeholder debates on just the key variables (e.g., the first 12\%).
% \item
%   There is  much evidence that many models   contain keys;
%   see \cite{Menzies2012,menzies2004many,druzdel94},  \tion{compsim}, \fig{softout} and \tab{nsga2_small}.
%   \ei
% Hence,
%We hope this work encourages others to check for  keys in their RE models.

% When not to use SHORT?
% We might recommend traditional goal model analysis (using forward and and backwards analysis)
% if the goal is only to increase user-acceptance of a single generated solution.
% However, SHORT is recommended when the objective is to
% \bi
% \item
% Assert the robustness of some solution {\em within the space
% of many others} or
% \item
% To show users a
%   ``trade space'' diagram (see \fig{neil1_star1})
% that lets then  decide when enough benefit has been obtained for enough cost.
% \ei
% Also, as mentioned above, SHORT could be used in combination with forward/backward analysis.

         %Runtimes increase polynominally on nodes
%edges nearly lunear
%- edges linear on nodes R^2 =0.93 e=49n - 21
%- runtime polynonimal on edges 7.6e^2 0 20e + 17 r^2 -0.91

\section*{Technical Appendix}
\subsection*{Reproduction Package}
All our tools and models are available on-line at \href{https://goo.gl/gvxeaH}{https://goo.gl/gvxeaH}. While some aspects
of our approach are specific to goal models, the general SHORT method could be applied
to a wide range of models.

\subsection*{Graph Smoothing}
To smooth out the charts generated in SHORT (e.g. Figure~\ref{fig:neil1_star1}), we use the 
     Scott-Knott procedure  recommended by \citep{mittas13}. This
     technique recursively bi-clusters a sorted
    set of numbers. If any two clusters are statistically indistinguishable, Scott-Knott
    reports them both as one  line.
   E.g. for lists $l,m,n$ of size $ls,ms,ns$ where $l=m\cup n$, Scott-Knott divides the sequence at the break that maximizes:
     \[E(\Delta)=\frac{ms}{ls}abs(m.\mu - l.\mu)^2 + \frac{ns}{ls}abs(n.\mu - l.\mu)^2\]
Scott-Knott then applies some statistical hypothesis test $H$ to check if $m,n$ are significantly different. If so, Scott-Knott then recurses on each division.
    For this study, our hypothesis test $H$ was a conjunction of the A12 effect size test (endorsed by
    \cite{arcuri11})  and non-parametric bootstrap sampling \cite{efron94}; i.e. our Scott-Knott divided the data if {\em both}
    bootstrapping and an effect size test agreed that the division was statistically significant (99\% confidence) and not a ``small'' effect ($A12 \ge 0.6$).
    
 \section*{Acknowledgments} % (fold)
\label{sec:acknow}
We would like to thank Dr. Jennifer Horkoff for providing us with these models
and for many insightful comments on earlier drafts of this paper.

This material is based upon work funded and supported by the Department of Defense under Contract No. FA8721-05-C-0003 with Carnegie Mellon University for the operation of the Software Engineering Institute, a federally funded research and development center. [Distribution Statement A] This material has been approved for public release and unlimited distribution. Please see Copyright notice for non-US Government use and distribution. DM-0004458

\footnotesize 
\bibliographystyle{IEEEtran}
\bibliography{modern}   % name your BibTeX data base
 
\end{document}

%% file: obj_reports.tex
\begin{table}
% \tiny
\footnotesize
\centering
\label{fig:softout}
\begin{tabular}{|l|l|l|l|l|l|l|l|}
\hline
\multirow{2}{*}{\textbf{Model}} & \multirow{2}{*}{\textbf{Result}} & \multicolumn{6}{l|}{\textbf{Percentage of Decisions}} \\ \cline{3-8}
 &  & \textbf{0} & \textbf{6} & \textbf{12} & \textbf{25} & \textbf{50} & \textbf{100} \\ \hline
\multirow{3}{*}{Counselling}  & SGs  & 60 & 60 & 70 & 80 & 70 & 60 \\
                              & Goals      & 50 & 60 & 70 & 60 & 70 & 40 \\
                              & Costs      & n.a. & 5 & 10 & 20 & 28 & 35 \\
                              \hline
\multirow{3}{*}{Management}   & SGs  & 50 & 50 & 60 & 60 & 50 & 40 \\
                              & Goals      & 50 & 50 & 50 & 60 & 60 & 60 \\
                              & Costs      & n.a. & 4 &  8 & 20 & 24 & 32 \\
                              \hline
\multirow{3}{*}{Marketing}    & SGs  & 70 & 70 & 70 & 60 & 60 & 30 \\
                              & Goals      & 70 & 70 & 70 & 70 & 60 & 40 \\
                              & Costs      & n.a. & 6 &  8 & 11 & 12 & 20 \\
                              \hline
\multirow{3}{*}{ITDepartment} & SGs  & 70 & 70 & 60 & 70 & 70 & 60 \\
                              & Goals      & 60 & 50 & 50 & 80 & 80 & 70 \\
                              & Costs      & n.a. & 2 &  4 &  9 & 17 & 18 \\
                              \hline
\multirow{3}{*}{SAProgram}    & SGs  & 80 & 80 & 80 & 80 & 80 & 80 \\
                              & Goals      & 70 & 70 & 70 & 60 & 55 & 60 \\
                              & Costs      & n.a. & 1 &  2 &  3 &  3 &  5 \\
                              \hline
\multirow{3}{*}{Services}     & SGs  & 50 & 50 & 50 & 50 & 50 & 50 \\
                              & Goals      & 30 & 50 & 60 & 60 & 70 & 70 \\
                              & Costs      & n.a. & 4 &  7 & 13 & 19 & 24 \\
                              \hline
\multirow{3}{*}{Kids\&Youth}  & SGs  & 50 & 40 & 50 & 50 & 40 & 20 \\
                              & Goals      & 50 & 50 & 70 & 60 & 60 & 70 \\
                              & Costs      &n.a. & 0 &  0  & 4 &  7 & 10 \\
                              \hline
\end{tabular}
\caption{Percentages of Softgoals and Goals covered, and total Costs, with respect to their maximum values after applying top 6,12,25,50,100\%
    of decisions found and ranked by SHORT for the
    goal models of \tref{tab:sizes}.}\label{tab:partials}
\end{table}